\documentstyle[preprint,12pt,aps,epsfig,psfig]{revtex}

\def\M{{\cal{M}}}

\def\bc{\begin{center}}
\def\ec{\end{center}}
\def\beq{\begin{equation}}
\def\eeq{\end{equation}}
\def\bea{\begin{eqnarray}}
\def\eea{\end{eqnarray}}
\def\bit{\begin{itemize}}
\def\eit{\end{itemize}}
\def\ben{\begin{enumerate}}
\def\een{\end{enumerate}}
\def\ba{\begin{array}}
\def\ea{\end{array}}
\def\bc{\begin{center}}
\def\ec{\end{center}}

\def\le{\left}
\def\r{\right}
\def\cd{\cdot}

\def\nl{\nonumber\\}

\def\ol{\overline}

\def\ga{\gamma}

\def\ep{\epsilon}

\def\th{\theta}

\def\dReg(#1,#2){\SetOffset(#1,#2)\BCirc(0,0){4}
    \Line(2.8,2.8)(-2.8,-2.8)
    \Line(2.8,-2.8)(-2.8,2.8)\SetOffset(0,0)}

\newcommand{\gl}[1]{{\!{\not\! #1}}}

\newenvironment{rowvec2}{\left(\begin{array}{cc}}{\end{array}\right)}
\newenvironment{colvec}{\left(\begin{array}{c}}{\end{array}\right)}
\def\bcol{\begin{colvec}}
\def\ecol{\end{colvec}}
\def\brow2{\begin{rowvec2}}
\def\erow2{\end{rowvec2}}

\begin{document}

\setlength{\parskip}{1.5ex plus 0.5ex minus 0.5ex}

\draft
\preprint{\begin{tabular}{l}
\hbox to\hsize{hep-ph/9804xxx \hfill SNUTP 98-027}\\[5mm] \end{tabular} }

\bigskip

\title{Analytic reduction of transition amplitudes in 
        radiative electroweak processes}
\author{Seungwon Baek\thanks{Electronic address: swbaek@phya.snu.ac.kr},
J.~S.~Shim\thanks
{Present address: Myongji University, Yongin, Korea.  jsshim@wh.myongji.ac.kr},
H.~S.~Song\thanks{Electronic address: hssong@physs.snu.ac.kr},
and Chaehyun Yu\thanks{
Electronic address: cyu@zoo.snu.ac.kr}
}
\address{
Center for Theoretical Physics and Department of Physics,\\ 
Seoul National University, Seoul 151-742, Korea 
}
\maketitle
\tighten
\begin{abstract}
It is shown that the transition amplitudes
of radiative electroweak processes like 
$e^-e^+ \rightarrow \nu_e \ol{\nu}_e \ga$,
$\ga e^- \rightarrow \nu_e \ol{\nu}_e e^-$,
and
$\nu_e e^- \rightarrow \nu_e e^- \ga$
can be reduced and factorized into simpler forms in the Standard Model.
This method can be used in reducing many calculations
of complicated radiative electroweak processes in
general.
\end{abstract} 

In our previous work~\cite{sbs95}, it has been shown that in the
Standard Model (SM) the 
$\ga e^- \rightarrow W^- \nu_e $ process can be treated mathematically
as a special case of the $\ga e^- \rightarrow Z^0 e^- $ process
which is described by the transition amplitude
\bea
 \M_{\ga e^- \rightarrow Z^0 e^-}
 &=&
 {e g_Z \over 2 } \ol{u}(p_2) \bigg[ 
 \frac{\gl{\ep}^*_Z (\gl{k}\,\gl{\ep}_\ga +2p_1\cd \ep_\ga)}{2k\cd p_1}
 + \frac{ (\gl{\ep}_\ga\, \gl{k} -2p_2\cd \ep_\ga)\gl{\ep}^*_Z}{2k\cd p_2}
 \bigg] \nl
 && \hspace{3cm} \times[\ep_L (1-\ga_5)+\ep_R (1+\ga_5)] u(p_1)\;.
\label{reze}
\eea
Here $k,p_1,$ and $p_2$ are the momenta of photon, incoming electron and
the outgoing electron, respectively. $\ep_\ga^\mu$
and $\ep_Z^\mu$ are the wave vectors of the photon and $Z^0$, and 
$g_Z,\ep_L$ and $\ep_R$ are coupling constants defined by
\bea
 g_Z = \le(8 m_W^2 G_F \over \sqrt{2} \cos^2 \th_W\r)^{1/2},
 \quad \ep_L = -{1\over2} +\sin^2\th_W,
 \quad \ep_R = \sin^2\th_W \;.
\eea
The $\ga W W$ coupling is included in the
$\ga e^- \rightarrow W^- \nu_e $ process, but  the transition amplitude
of the $\ga e^- \rightarrow W^- \nu_e $ process can be changed in the
form of Eq.~(\ref{reze}) due to the following relation
\bea
 &&\frac{\gl{\ep}_W^* (\gl{k}\,\gl{\ep}_\ga + 2 p_1 \cd \ep_\ga)}{k\cd p_1}
 + {2 \over k\cd Q} [\ep_\ga \cd \ep_W^* \gl{k} -k\cd \ep_W^* \gl{\ep}_\ga^*
                     -\ep_\ga \cd Q \gl{\ep}_W^*] \nl
 && \hspace{1cm} = -{k\cd p_2 \over k \cd Q}
   \le[\frac{\gl{\ep}_W^* (\gl{k}\,\gl{\ep}_\ga + 2 p_1 \cd \ep_\ga)}{k\cd p_1}
   +\frac{ (\gl{\ep}_\ga \,\gl{k} - 2 p_2 \cd \ep_\ga) \gl{\ep}_W^*}{k\cd p_2}
   \r],
\label{tran}
\eea
where $\ep_W^\mu$ and $Q^\mu$ are the wave vector and momentum of
$W^-$ and $p_2$ in Eq.~(\ref{tran}) implies the momentum of the outgoing
neutrino. Therefore, various results obtained in the 
$\ga e^- \rightarrow Z^0 e^- $ process can be used in the 
$\ga e^- \rightarrow W^- \nu_e $ process after the coupling 
constants $\ep_R,\ep_L g_Z$
and mass $m_Z$ in the former process are replaced by
$0, -g_W (k\cd p_2 / k\cd Q)$ and $m_W$ in the  latter process.

In the SM, the processes 
$e^-e^+ \rightarrow \nu_e \ol{\nu}_e \ga$,
$\ga e^- \rightarrow \nu_e \ol{\nu}_e e^-$
and
$\nu_e e^- \rightarrow \nu_e e^- \ga$
can be described by five Feynman diagrams in the tree level~\cite{comphep}.
They consist of two diagrams related with the $Z^0$-mediated process,
one diagram with the $\ga W W$ coupling and two diagrams
involving the $W^-$ particle as shown in Fig.~1 for the
$e^-e^+ \rightarrow \nu_e \ol{\nu}_e \ga$~\cite{nu3}.
Two terms in the transition amplitudes involving the $W^-$ particle
can be changed into the terms corresponding to the first two diagrams
by the Fierz transformation, and using the identity of Eq.~(\ref{tran}),
the $\ga W W$ term in the transition amplitude can be resolved into the
terms of the first two diagrams similarly. This implies that the transition
amplitudes of the process 
$e^-e^+ \rightarrow \nu_e \ol{\nu}_e \ga$,
$\ga e^- \rightarrow \nu_e \ol{\nu}_e e^-$
and
$\nu_e e^- \rightarrow \nu_e e^- \ga$ can be essentially
described by the expression for two diagrams which involve the 
$\ga e^- \rightarrow Z^0 e^- $ process after proper modification of
$(1-\ga_5)$ term in the processes. Therefore, 
we expect that the total transition amplitude can be factorized 
due to two neutrinos coming from the decay $Z^0 \rightarrow \nu_e \ol{\nu}_e$.

Explicitly the transition amplitude of the process
$e^-(p_1)e^+(p_2) \rightarrow \nu_e(k_1) \ol{\nu}_e(k_2) \ga(k)$
can be obtained as follows,
{
\setcounter{enumi}{\value{equation}}
\addtocounter{enumi}{1}
\setcounter{equation}{-1}
\renewcommand{\theequation}{\theenumi\alph{equation}}
\bea
 \M_{e^-e^+ \rightarrow \nu_e \ol{\nu}_e \ga} 
 &=& {e g_W^2 \over 8 \cos^2\th_W}{1 \over (2 k_1\cd k_2 -m_Z^2)} \nl
     &&\hspace{-1.2cm}\times\ol{v}(p_2) 
     \bigg[ \frac{\ga^\mu (\gl{\ep}_\ga^* \, \gl{k} + 2 p_1\cd \ep_\ga^*)}
                 {2 k\cd p_1}
		 \big[E_{1L} (1-\ga_5) + \ep_R (1+\ga_5) \big]  \nl
             &&+ \frac{(\gl{\ep}_\ga^* \, \gl{k} - 2 p_2\cd \ep_\ga^*)\ga^\mu }
                 {2 k\cd p_2}
                 \big[E_{2L} (1-\ga_5) + \ep_R (1+\ga_5) \big] \nl
            && + \ga^\mu (k_2 -p_1)\cd \ep_\ga^* A_L (1-\ga_5) \bigg] u(p_1) 
           \ol{u}(k_1) \ga_\mu (1-\ga_5) v(k_2),
\label{eennr}
\eea
where
\bea
E_{1L} &=& \ep_L 
          +\cos^2 \th_W  
	   \frac{(2 k_1\cd k_2 -m_Z^2)[2 p_1\cd(k +k_1) +m_W^2 ]}
	        {(2 p_1\cd k_1 +m_W^2)(2 p_2 \cd k_2 +m_W^2)} ,\\
E_{2L} &=& \ep_L 
          + \cos^2 \th_W 
           \frac{(2 k_1\cd k_2 -m_Z^2)[2 p_2\cd(k +k_2) +m_W^2 ]}
                {(2 p_1\cd k_1 +m_W^2)(2 p_2 \cd k_2 +m_W^2)} , \\
A_L    &=&  2 \cos^2 \th_W
            \frac{(2 k_1 \cd k_2 -m_Z^2)}
                 {(2 p_1 \cd k_1 +m_W^2)(2p_2\cd k_2 +m_W^2)}.
\eea
\setcounter{equation}{\value{enumi}}
}
\hspace{-0.38cm} The third term 
in the square bracket of Eq.~(\ref{eennr}) is necessary
for the transition amplitude to have gauge invariance after the 
$\ep_L (1-\ga_5)$ term is modified. 

The transition amplitude $\M_{\ga e\rightarrow \nu_e\ol{\nu}_e e}$
for the $\ga(k) e^-(p_1)\rightarrow \nu_e(k_1)\ol{\nu}_e(k_2) e^-(p_2)$
process can be obtained from
Eq.~(\ref{eennr}) for $\M_{e^-e^+ \rightarrow \nu_e \ol{\nu}_e \ga}$ 
by replacing $k,p_2,{v}(p_2)$, and $\ep_\ga^*$ by
$-k,-p_2,{u}(p_2)$, and $\ep_\ga$, respectively.
Also the transition amplitude $\M_{\nu_e e\rightarrow \nu_e e \ga}$
for the $\nu_e(k_2) e(p_1)\rightarrow \nu_e(k_1) e(p_2) \ga(k)$
can be obtained similarly from Eq.~(\ref{eennr}) of 
$\M_{e^-e^+ \rightarrow \nu_e \ol{\nu}_e \ga}$ by replacing
$k_2, p_2, v(k_2)$, and $v(p_2)$ by $-k_2, -p_2, u(k_2)$, and  $u(p_2)$,
respectively.  These processes can be described essentially
by Fig.~2 (a)-(c).

One can see that, in the
$e^- e^+ \rightarrow \nu_e \ol{\nu}_e \ga$ 
process, if the initial
energy is about 100 GeV, the contribution from the 
$Z^0$-mediated diagrams of Fig.~1 is dominant and as the incident 
energy increases, the
contribution from other diagrams becomes large. 
The polarization effect in 
$e^- e^+ \rightarrow Z^0 \ga \rightarrow f \ol{f} \ga$ has
been considered by us~\cite{song} and the method can be applied here.

\acknowledgements

The authors would like to thank S.~Y.~Choi and S.~B.~Kim for useful
discussion. The work is supported in part by KOSEF through the SRC
program and in part by the Korean Ministry of Education through the
BSR program, BSRI-97-2418.



\newpage
\begin{center}
{\Large\bf FIGURE CAPTIONS}
\end{center}
\noindent Fig.~1,
Feynman diagrams for the process 
$e^- e^+ \rightarrow \nu_e \ol{\nu}_e \ga$.
\\
\noindent Fig.~2,
Reduced Feynman diagrams for 
(a) $e^- e^+ \rightarrow \nu_e \ol{\nu}_e \ga$,
(b) $\ga e^- \rightarrow e^- \nu_e \ol{\nu}_e $,
(c) $\nu_e e^- \rightarrow \nu_e e^- \ga $.
\vskip1cm

\newpage
\thispagestyle{empty}
\begin{figure}[htb]
 \unitlength 1mm
\begin{picture}(100,100)
  \put(23,-50){ \epsfxsize=14cm \epsfysize=18cm \epsfbox{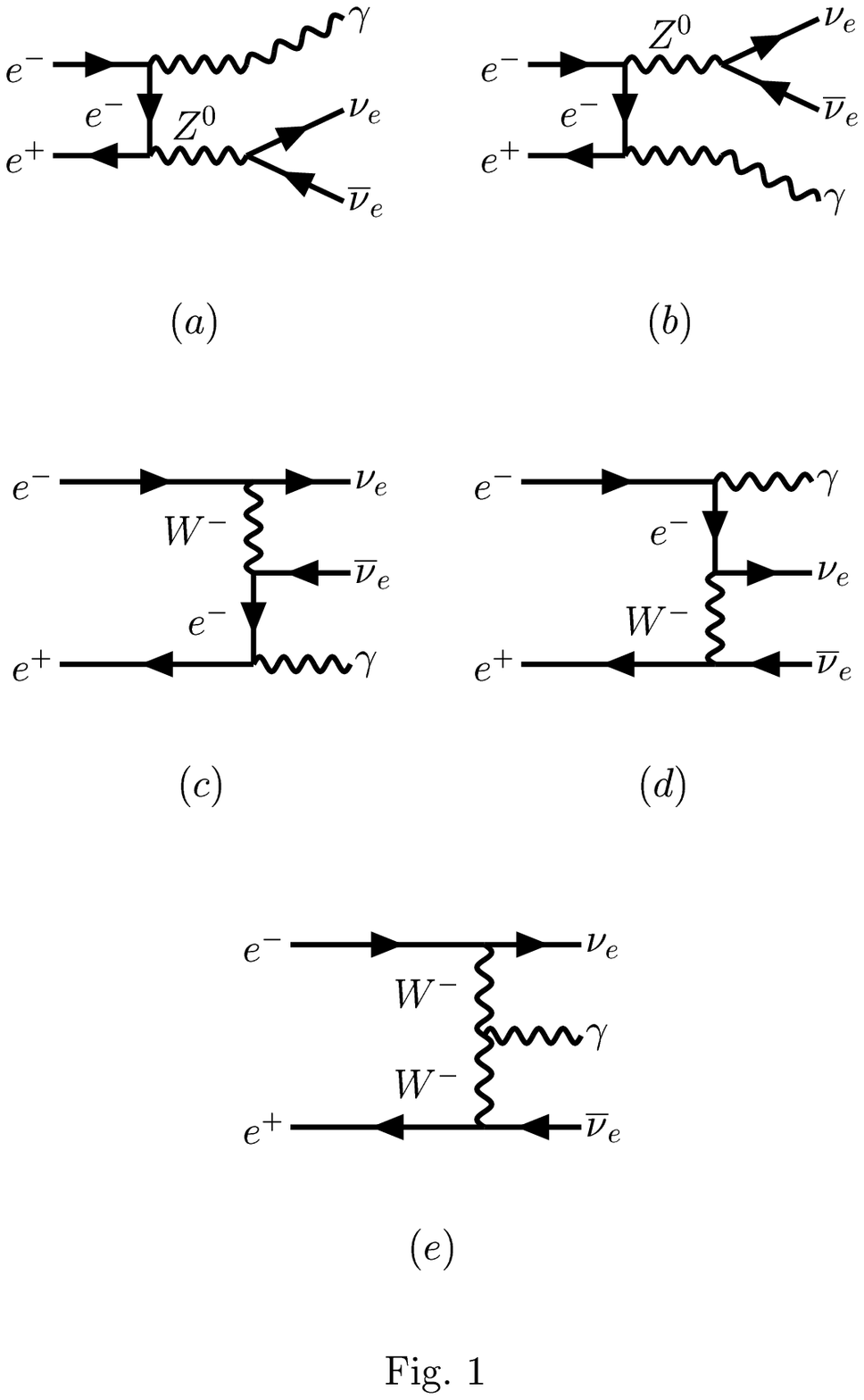}}
\end{picture}
\end{figure}

\begin{figure}[htb]
 \unitlength 1mm
\begin{picture}(100,100)
  \put(12,-10){ \epsfxsize=15.5cm \epsfysize=6.5cm \epsfbox{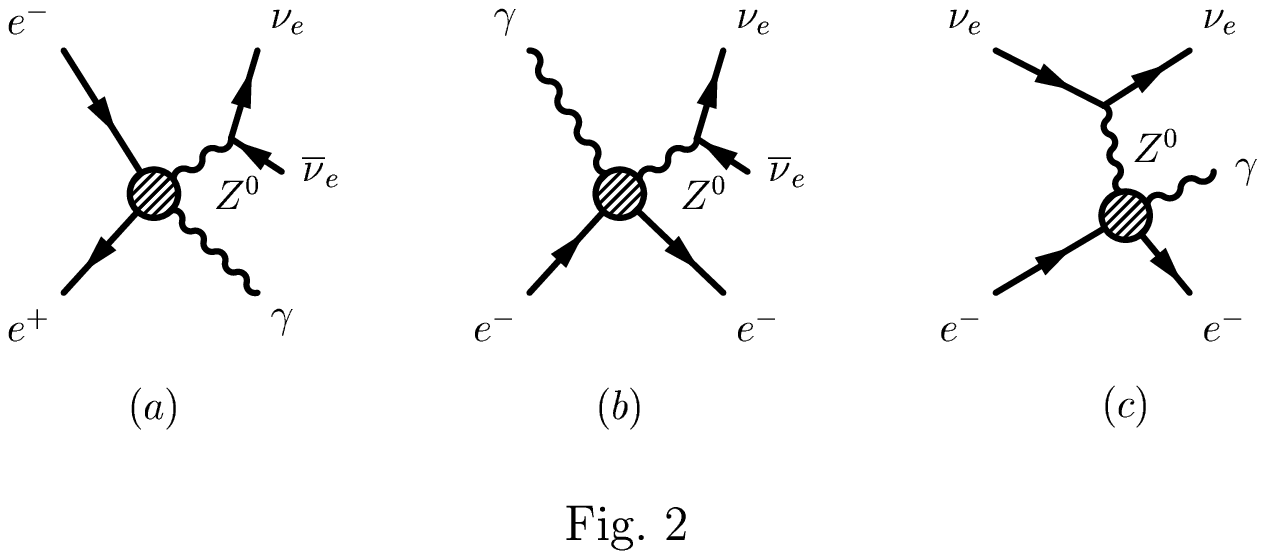}}
\end{picture}
\end{figure}

\end{document}